%% file: main.tex
\documentclass[10pt,letterpaper,compsoc,conference]{iiswc26}

\usepackage{cite}
\usepackage{amsmath,amssymb,amsfonts}
\usepackage{algorithmic}
\usepackage{graphicx}
\usepackage[dvipsnames]{xcolor}
\usepackage[final]{microtype}
\usepackage[italic]{mathastext}
\usepackage{libertine}
\usepackage[T1]{fontenc}
\usepackage{textcomp}
\usepackage[varqu,varl]{zi4}
\usepackage[all]{nowidow}
\usepackage[keeplastbox]{flushend}
\usepackage{fancyhdr}
\usepackage{booktabs}
\usepackage{tikz}
\usepackage{graphicx}
\usepackage{xcolor}   
\usepackage{multirow}
\usepackage{makecell}
\usepackage{svg}
\usetikzlibrary{arrows.meta, positioning, fit, backgrounds, calc}
\usepackage[switch]{lineno}
\usepackage{longtable, array, xcolor, booktabs}

\usepackage[normalem]{ulem} 

\fancypagestyle{firstpage}{
  \fancyhf{}
  
  \fancyfoot[C]{\thepage}
}

\begin{document}


\title{Characterizing LLM Kernel Access and Memory Interaction in Multi-Partition NUMA GPUs}





\author{%
Donghyeon Joo\textsuperscript{*1,2}, Sooraj Puthoor\textsuperscript{2}, Nuwan Jayasena\textsuperscript{2}, and Bahar Asgari\textsuperscript{1}\\[3pt]
\textsuperscript{1}University of Maryland, College Park \quad \textsuperscript{2}AMD\\[3pt]
{\normalfont\ttfamily \{dhjoo98,\,bahar\}@umd.edu,\ \{sooraj.puthoor,\,Nuwan.Jayasena\}@amd.com}%
}

\maketitle
\begingroup
\renewcommand{\thefootnote}{}
\footnotetext{\hspace{-1.8em}\textsuperscript{*}Work done during internship at AMD.}
\endgroup
\thispagestyle{firstpage}
\pagestyle{plain}


\begin{abstract} Large language model (LLM) workloads motivate multi-partition GPUs as a path to scaling compute and memory capacity, but their non-uniform memory access characteristics and inter-partition communication can amplify contention and degrade locality, leading to suboptimal kernel latency. To address this, we analyze performance-critical LLM kernel implementations spanning weight projection, mixture-of-experts, and attention variants of state-of-the-art serving engines to present a characterization of data access patterns in multi-partition GPUs.
First, we introduce memory trace analysis methodology to derive workgroup-level data access and sharing behavior, then evaluate the locality implications on latency using a cycle-level simulator.
Using these tools, we categorize LLM kernel operands into three inter-workgroup sharing patterns (global, partial, or private) and show that the required optimization strategies differ across categories, from simple per-workgroup pinning to subgroup-aware co-scheduling.
Our findings highlight the need for placement-aware kernel programming and smarter architectural support for work and data locality in multi-partition GPUs.
\end{abstract}

\input{src/01_introduction}
\input{src/02_backgrounds}
\input{src/03_methodology}
\input{src/04_categorization}

\input{src/05_evaluation}

\section{Conclusions}
We characterize memory access patterns of performance-critical LLM kernels on multi-partition NUMA GPUs, classifying them by inter-workgroup sharing pattern (global, partial, private). 
Our trace-driven analysis and cycle-level simulation reveal that state-of-the-art  LLM serving kernels have significant opportunities for speedup through careful placement of workgroups and data structure slices across partitions, and that current runtimes cannot fully exploit these opportunities.
These findings underscore the need for placement-aware WG scheduling, finer data interleaving granularity, and runtime support for partition locality to fully realize the memory capacity benefits of multi-partition GPU architectures.



\bibliographystyle{IEEEtranS}
\bibliography{reference}

\end{document}

%% file: src/01_introduction.tex
\section{Introduction}
\label{sec:introduction}

Large language models (LLMs) have grown rapidly in both parameter count and deployment scale, placing mounting pressure on accelerator platforms to deliver higher compute throughput and larger effective memory capacity simultaneously. Models such as DeepSeek~\cite{deepseekv2}, Gemma~\cite{GemmaTeam2025}, and Qwen~\cite{Xuetetal2025} now routinely exceed hundreds of billions of parameters, and inference workloads increasingly demand long context windows, pushing effective KV cache footprints well beyond what a single memory stack can serve efficiently. At the same time, architectural trends such as mixture-of-experts (MoE) and sparse attention have decoupled model capacity from per-token compute, introducing irregular sparse activation patterns that further stress memory subsystems.

On the serving side, inference engines such as vLLM~\cite{kwon2023vllm} and SGLang~\cite{zheng2024sglang} have emerged to 
expose a rich kernel-level interface through which memory access patterns are ultimately determined, patterns whose locality properties are becoming increasingly consequential as hardware scales.

To meet the capacity and bandwidth demands of LLM inference, GPU vendors are turning to multi-partition designs. Rather than scaling a monolithic die, multi-partition GPUs compose several compute and memory domains within a single package, connected by a high-bandwidth on-package interconnect. AMD Instinct\texttrademark~MI300X~\cite{amdmi300x2025} and NVIDIA B200~\cite{jarmusch2025blackwell} are early examples of this integration strategy~\cite{lohNextEraChiplet2023, xuExploringWaferScaleGPUs2025} in the industry, with parallel efforts in academia exploring multi-chiplet composition for DNN accelerators to address analogous scaling and communication challenges~\cite{shao2019simba, cai2024gemini, tan2021nnbaton, musavi2025communication}.
However, this composition comes at the cost of non-uniform memory access (NUMA). A workgroup executing on one partition may access data resident on a remote partition, incurring higher latency and competing for shared interconnect bandwidth. Non-uniform memory access effects now manifest at intra-package level.

Current GPU memory management units (MMUs) expose a single unified virtual address space to kernels, abstracting away the physical partition layout. While this simplifies programming, it also means neither the kernel developer nor the runtime has explicit fine-grained control over where data physically resides relative to the workgroups that consume it. The consequence is that access patterns which are benign on monolithic GPUs, shared weight tensors, activations, or KV cache entries, can silently become cross-partition traffic, amplifying contention and degrading latency on multi-partition platforms. 
To this end, we ask \textit{what are the costs of non-uniform memory access in modern LLM serving kernels, and what kernel programming practice and architectural support are required to close the gap?}

To answer this, we conduct a systematic characterization of LLM kernel memory access patterns through the lens of multi-partition NUMA effects. We analyze performance-critical operators, weight projection, attention variants, and mixture-of-experts, as implemented in vLLM and SGLang by first generating memory traces from real kernel execution.   
We further develop GPU memory trace analysis techniques to derive inter-workgroup sharing behavior of target kernels to classify kernels by their \textit{inter-workgroup sharing patterns} (global, partial, or private).
Second, we extend an existing cycle-level simulator that consumes target kernel objects to analyze the NUMA implications on latency.
Together, these methods expose the structural properties that determine NUMA sensitivity and guide both architectural design and kernel development practice.







%% file: src/02_backgrounds.tex
\section{Background}
\label{sec:background}

This section provides background on multi-partition NUMA GPUs, how the MMU exposes the address space, and the LLM workload scope we target.

\subsection{Multi-partition GPU}

\begin{figure}[h]
   \centering
   \includegraphics[width=\linewidth]{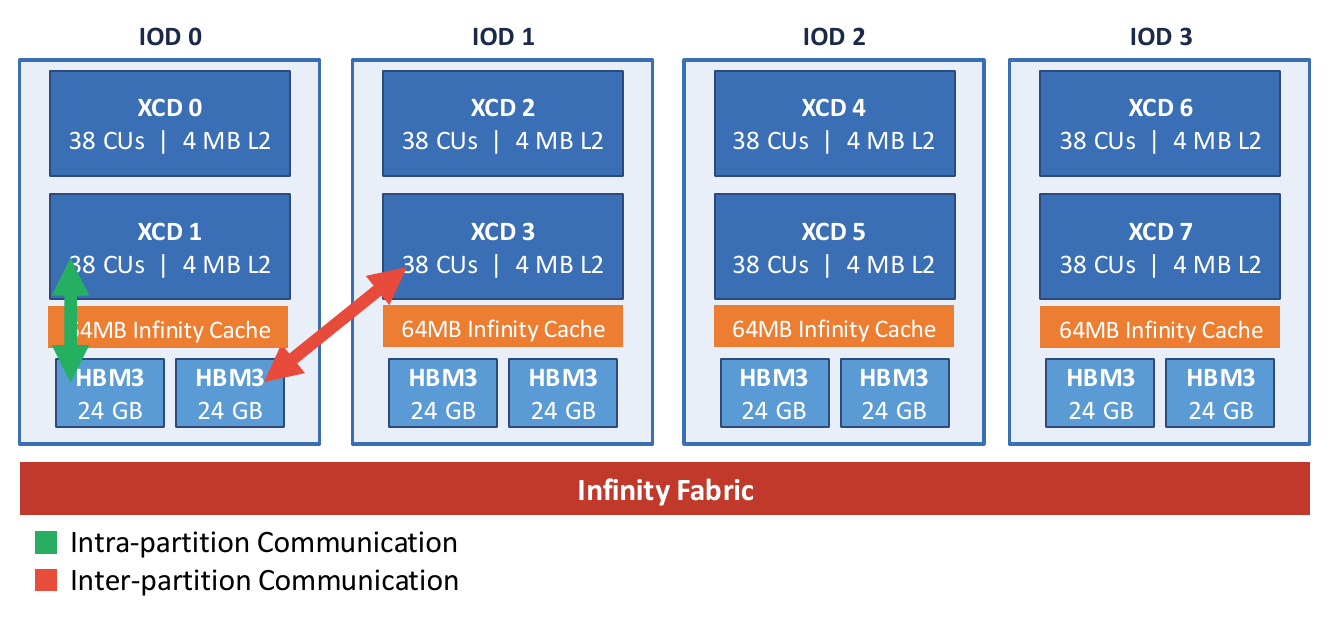}
   \caption{Multi-Partition Architecture of AMD Instinct MI300X.}
   \label{fig:mi300x}
 \end{figure}

AMD Instinct MI300X (Figure~\ref{fig:mi300x})~\cite{smithRealizingAMDExascale2024} is a prominent example of how GPU compute and memory scaling is achieved through a multi-partition architecture~\cite{arunkumarMCMGPUMultiChipModuleGPUs2017, milicSocketNUMAawareGPUs2017}. The package integrates eight Accelerator Compute Dies (XCDs), each containing 38 Compute Units (CUs) and a 4\,MB private L2 cache, for a total of 304 CUs and 32\,MB of aggregate L2. The eight XCDs are organized as four pairs, where each pair is 3D-stacked on top of a dedicated I/O die (IOD). The four IODs provide on-package networking and connectivity to eight HBM3 stacks that together supply 192\,GB of memory~\cite{choudharyOptimizingAttentionGPUs2025}. 
A 256\,MB Infinity Cache, distributed across the four IODs, acts as a last-level cache between the per-XCD L2 caches and HBM3, reducing the frequency of accesses that must reach external memory. 
The XCDs, IODs, and HBM stacks are interconnected via AMD's Infinity Fabric, a high-bandwidth on-package interconnect that routes data requests across the chip complex, while Infinity Cache intercepts a fraction of those requests on the IOD before they reach HBM~\cite{amdmi300x2025}.

When a workgroup scheduled on a CU accesses data that is mapped to the HBM of the same IOD, data access overhead (green arrow of Figure~\ref{fig:mi300x}) is minimal compared to accessing data mapped to HBM of a remote IOD.
Inter-partition accesses (red arrow of Figure~\ref{fig:mi300x}) must traverse the interconnect and thus exhibit substantially higher latency. 
Therefore, the natural abstraction required for locality-aware scheduling and data placement is the \emph{base die}, which is the physical die whose compute dies share local HBM. On the AMD Instinct MI300X, each IOD and two XCDs constitute one base die. 
Each base die forms a single partition, the scheduling and data-placement unit we use throughout this paper, as the same abstraction applies to any multi-partition GPU in which a group of compute units shares a local memory domain.

\paragraph{\textbf{The Abstraction Gap}}
Despite this physical NUMA structure, the GPU memory management unit (MMU) exposes a single, unified virtual address space to all kernels and programmers. This abstraction simplifies programming but critically hides the partition topology where neither the kernel developer nor the runtime have explicit, fine-grained visibility into where data physically resides relative to the workgroups that consume it. 
The physical placement of compute and data across partitions, which determines whether a given access is served locally or must traverse the partition boundary, is abstracted away at the programming level. As a consequence, access patterns that are benign on monolithic GPUs, such as shared weight tensors, activations, or reused KV cache entries, can silently become cross-partition traffic on multi-partition platforms, amplifying contention and degrading kernel latency without any indication to the programmer. This tension has been identified across interconnect and cache architecture~\cite{milicSocketNUMAawareGPUs2017, kimCODAEnablingColocation2018, power2013hsc}, virtual memory management~\cite{pratheekDesigningVirtualMemory2022, tanACOPTAdaptiveContinuityaware2026, fengBarreChordEfficient2024}, memory migration~\cite{prodromou2017mempod, bakhtiar2025chason}, and inter-partition synchronization~\cite{zhong2026lrmgpu}, and is the central problem our work addresses.

\subsection{Prior Characterization Work}
\label{sec:related_work}

\paragraph{\textbf{LLM Inference Characterization}}
A growing body of work characterizes LLM inference at the kernel and system levels. Splitwise~\cite{patel2024splitwise} decomposes inference into prefill and decode phases and proposes phase-aware scheduling to reduce interference. Subsequent studies characterize memory-subsystem bottlenecks under large-batch inference~\cite{recasens2025memorygap}, heterogeneous CPU--GPU data-movement overheads~\cite{vellaisamyCharacterizingOptimizingLLM2025}, compute--memory trade-offs across LLM kernels and model families~\cite{wangSystematicCharacterizationLLM2025}, and kernel-architecture co-design for sparse LLM inference~\cite{joo2025coruscant}. Collectively, these works provide valuable insights into system-level and kernel-level behavior during LLM inference. However, they do not examine the \emph{intra-package} NUMA effects that emerge when workgroups from a single kernel are distributed across partitions of a multi-chiplet GPU. Our work complements these efforts by characterizing memory access behavior at workgroup granularity and quantifying the impact of partition locality.

\paragraph{\textbf{Inter-Thread-Block Locality and Co-Scheduling}}
Exploiting data sharing across thread blocks to guide scheduling has a long history on monolithic GPUs. Prior work constructs inter-CTA locality graphs to co-schedule thread blocks with shared data footprints on the same SMs, improving cache locality and reducing redundant memory traffic~\cite{li2017locality}. PAVER~\cite{tripathy2021paver} extends this approach using compile-time PTX analysis and graph partitioning to maximize thread-block locality, while Huzaifa et~al.~\cite{huzaifa2020interkernel} study inter-kernel producer--consumer reuse across kernel boundaries. 
However, they target monolithic GPUs, where suboptimal placement primarily increases cache misses. In contrast, our work studies locality in multi-partition NUMA GPUs, where misplaced accesses incur expensive inter-partition communication.


%% file: src/03_methodology.tex
\section{Methodology}
\label{sec:methodology}
Our method starts from generating memory traces in the virtual address space from real GPU kernels. From these memory traces, we use our analysis logic to diagnose workgroup-level access patterns. 
We then quantify the performance impact of these patterns using cycle-level simulation, whose configuration we describe alongside the case-by-case results in Section~\ref{sec:setup}.

\noindent{\textit{\textbf{Why Workgroup-level?}}}
We conduct our analysis at the granularity of individual workgroups (equivalently, Threadblocks in NVIDIA terminology) because the workgroup is the fundamental unit of scheduling across Compute Units (CUs) in the AMD GPU execution model. In a multi-partition GPU such as the AMD Instinct MI300X, the hardware scheduler distributes workgroups across XCDs, interleaving them across partitions at dispatch time. Consequently, the memory access footprint and sharing behavior of a workgroup directly determines whether its accesses are satisfied locally within the partition's memory hierarchy or must traverse the inter-partition interconnect to reach a remote HBM stack. Analyzing at a coarser granularity (e.g., the full kernel grid) would obscure these per-partition locality effects, while analyzing at a finer granularity (e.g., individual wavefronts or lanes) would not align with the scheduling boundary that governs data placement decisions. Thus, characterizing which data each workgroup reads and writes, how workgroups share data with one another, and how their footprints relate spatially to partition boundaries is an adequate level of abstraction for reasoning about NUMA effects in multi-partition GPU execution.

\paragraph{\textbf{Virtual Address to Physical Address Mapping}} We use AMD Omniprobe~\cite{amdresearch_omniprobe} to collect kernel memory traces. Omniprobe trace collection operates in the virtual address space, as it instruments kernel execution at the application level where only virtual addresses are visible. A natural concern is whether virtual-address traces faithfully represent the partition-level locality effects that arise from physical page placement.
We argue that they do, as the structural properties we characterize, which workgroups share data and how many bytes they share, are invariant under any virtual-to-physical address mapping. Whether inter-WG overlap is zero, partial, or total is determined by the kernel's access pattern, not by where the runtime places pages. Our categorization (Section~\ref{sec:workload_characterization}) is therefore grounded in virtual-address arithmetic without loss of generality.



\subsection{Memory Trace Analysis}
This section describes the memory trace analysis pipeline that extracts inter-workgroup sharing behavior from raw GPU execution traces.
 
\subsubsection{Trace Ingestion and Address Filtering}
\label{sec:trace-ingestion}

\begin{figure}[h]
    \centering
    \includegraphics[width=\linewidth]{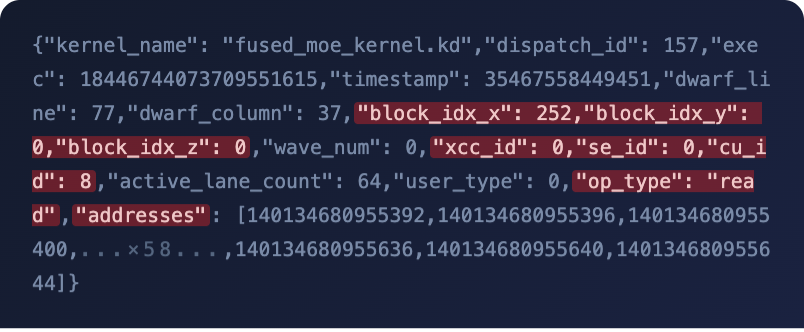}
    \caption{Example of Omniprobe Memory Trace.}
    \label{fig:trace_example}
\end{figure}

The pipeline ingests GPU execution traces stored as JSON files. As shown in Figure~\ref{fig:trace_example}, each trace entry carries the kernel name, dispatch ID, workgroup coordinates, wave number, operation type, and a list of per-lane memory addresses. We filter out entries whose addresses fall outside the global memory range. All subsequent analysis operates exclusively on the retained entries.
 
 

\subsubsection{Kernel and Workgroup Selection}
\label{sec:kernel-wg-selection}
 
From the filtered trace, we enumerate all unique kernel dispatches (identified by dispatch ID and kernel name) and allow the user to select a target dispatch. Within that dispatch, we enumerate all unique workgroup coordinates and extract the trace subset belonging to a chosen reference WG for single-WG analysis.
 
\subsubsection{Intra-Lane Stride Detection}
\label{sec:intra-lane-stride}
 
 
Before per-entry analysis, we characterize the intra-lane stride
$\sigma_{\mathrm{lane}}$, defined as the step size at which a single
wavefront lane advances through memory across successive instructions,
by collecting the first address per trace entry for a fixed wave,
segmenting the sorted sequence into constant-stride runs, and taking
the dominant stride. This stride is then used to determine byte-range
endpoints during chunk formation.
 
\subsubsection{Per-Entry Multi-Stride Segmentation}
\label{sec:per-entry-stride}
 
Each trace entry contains a list of addresses, one per active lane in a wavefront. We walk the consecutive deltas within that list and partition the sequence into stride-consistent sub-sequences. Specifically, the first non-zero delta initializes a \emph{base stride}. Subsequent deltas that match the base stride extend the current segment, while a mismatch closes the current segment and resets the base stride.  Zero-delta addresses (broadcast accesses, where multiple lanes target the same location) are recorded separately and counted as stride-zero events.
 
 
 
 
\subsubsection{Chunk Formation and Merging}
\label{sec:chunk-formation}
 
Stride-consistent segments are converted to byte intervals $[s,\;e)$ where $s = \min(\mathrm{sublist})$ and $e = \max(\mathrm{sublist}) + \sigma$.  All intervals sharing the same stride are then merged greedily. Two intervals that are contiguous or within a configurable gap tolerance $\delta$ are collapsed into a single interval.
The result is a compact set of \emph{merged memory chunks} per stride per operation type, representing the effective memory footprint of the workgroup for that stride class.
 
\subsubsection{Inter-workgroup Reuse Analysis}
\label{sec:inter-wg-reuse}
 
Chunk derivation is executed independently for every WG in a kernel. The resulting chunk sets are compared pairwise against a reference WG.  For a reference WG $R$ and another WG $Q$, we compute the byte-level overlap between corresponding chunk sets:
 
{\small
\begin{align*}
  \text{reuse}_{\mathrm{read}}(R, Q)
    &= \sum_{r \in \mathcal{C}^{\mathrm{rd}}_R}
       \sum_{q \in \mathcal{C}^{\mathrm{rd}}_Q}
       \max\!\bigl(0,\;\min(r_e, q_e) - \max(r_s, q_s)\bigr)
  \\
  \text{reuse}_{\mathrm{write}}(R, Q)
    &= \sum_{r \in \mathcal{C}^{\mathrm{wr}}_R}
       \sum_{q \in \mathcal{C}^{\mathrm{wr}}_Q}
       \max\!\bigl(0,\;\min(r_e, q_e) - \max(r_s, q_s)\bigr)
\end{align*}
}
 
\noindent where $\mathcal{C}^{\mathrm{rd}}_W$ and $\mathcal{C}^{\mathrm{wr}}_W$ denote the read and write chunk sets of WG $W$, and $(r_s, r_e)$ denotes the start and end of an interval.  Non-zero overlap quantifies shared data and is a direct proxy for potential NUMA contention when WGs are placed on different partitions.

%% file: src/04_categorization.tex
\section{Workload Categorization}
\label{sec:workload_characterization}

We apply the memory trace analysis methodology of Section~\ref{sec:methodology} to classify the memory access behavior of LLM kernels.
We characterize the memory access behavior of LLM kernels by analyzing how workgroups within a single dispatch collectively access each operand data structure. 

\subsection{Inter-WG Sharing Patterns}
\label{sec:intra-kernel}

\begin{figure}[h]
    \centering
    \includegraphics[width=\linewidth]{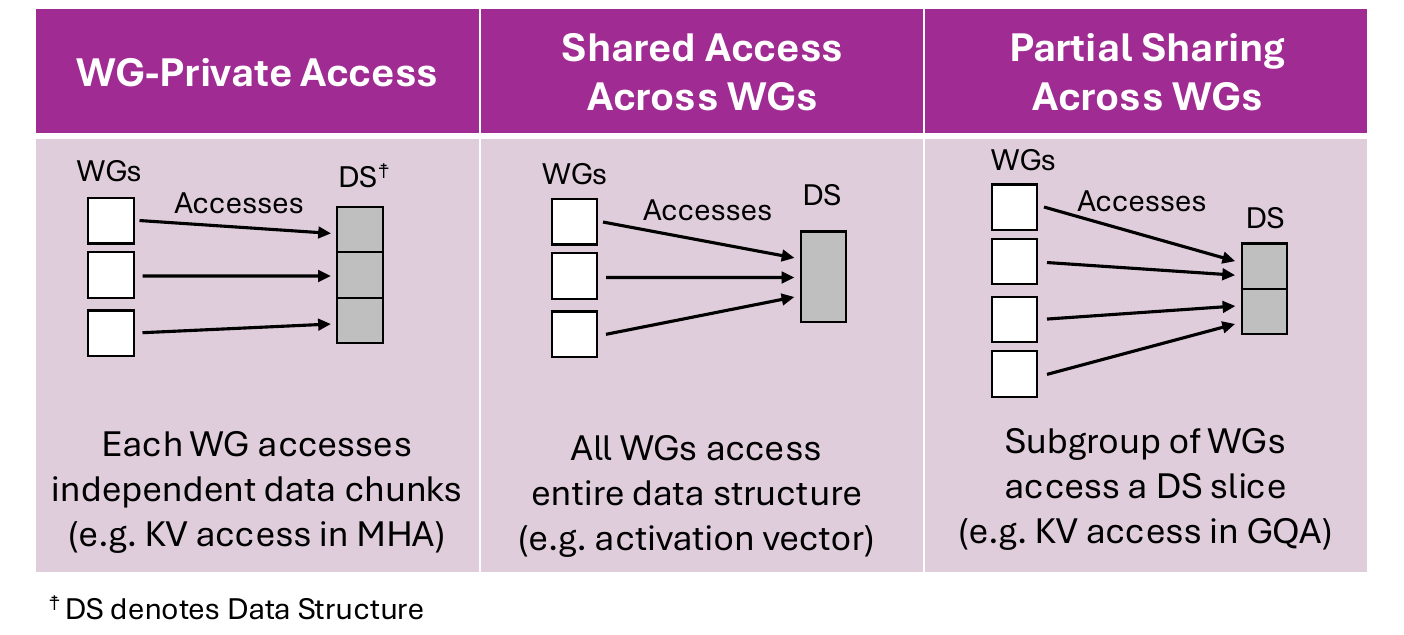}
    \caption{Categorization of WG-level Data Structure Access.}
    \label{fig:categorization}
\end{figure}

We identify three distinct patterns by which workgroups access a shared data structure. As shown in Figure~\ref{fig:categorization}, they are WG-local, global shared, and partial sharing across WGs. 

\paragraph{\textbf{WG-Private Access}}
Each workgroup reads or writes a disjoint region of the operand. There is zero measured overlap between the chunk sets of any two WGs. The operand is effectively \emph{partitioned} across the WG grid, and no inter-WG data sharing occurs within the kernel.
This pattern is locality-friendly since if the runtime can co-locate a WG with the memory region it exclusively owns, all accesses will be partition-local with no cross-partition replication.
Examples include the output tile of a GEMM-based projection kernel, where each WG is responsible for computing and writing a distinct output submatrix. 
 
\paragraph{\textbf{Global Shared Access}}
All workgroups in the dispatch access the entire extent of the operand, where our analysis reports large byte-overlap between every
WG pair that matches the full size of the structure.
The operand is \emph{broadcast} across the WG grid, in that every partition that receives WGs must either cache a local copy or incur repeated remote fetches.
We identify this pattern as the primary source of cross-partition contention in LLM kernels.
The clearest instance is the activation in a linear projection where regardless of which output tile a WG is computing, it must read across the full vector, meaning all WGs on all partitions compete for the same underlying data.
 
\paragraph{\textbf{Partial Sharing Across WGs}}
A structurally intermediate pattern in which workgroups form subgroups: workgroups within a subgroup share a slice of the operand, while workgroups in different subgroups access disjoint slices.
The byte-overlap measured between WGs is non-zero but bounded to a strict fraction of the operand.
This is the most nuanced pattern from a placement perspective since the relevant unit of co-location is the subgroup, not the individual workgroups and not the entire grid.
Examples include grouped-query attention~(GQA), where multiple query-head workgroups share the same K and V head, and mixture-of-experts~(MOE), where multiple expert-specific workgroups share the expert weight and activation vectors. 

\subsection{NUMA Sensitivity: Partition Locality}
\label{sec:partition-locality}

The categorization above describes how workgroups interact with each operand, but does not directly quantify the resulting cross-partition traffic. Therefore, we introduce a theoretical per-operand metric that captures the fraction of memory accesses served by each partition's local memory under a given placement and scheduling policy.

For an operand~$\mathcal{O}$ accessed by the workgroups of a kernel dispatch under placement policy~$\pi$ and workgroup scheduling policy~$\sigma$, we define Partition Locality as
{\small
\begin{equation*}
  \mathrm{Partition\;Locality}(\mathcal{O},\pi,\sigma)
  \;=\;
  \frac{\#\text{local\_accesses}}
       {\#\text{local\_accesses} + \#\text{remote\_accesses}}
\end{equation*}
}
%

\paragraph{\textbf{Decomposition}}
Partition Locality is determined by the composition of three independently controllable factors:
\begin{enumerate}
  \item 
\textbf{Kernel access footprint.}  The per-operand virtual address range each workgroup touches and the degree of overlap across WGs.  This is extracted by our trace analysis.

  \item \textbf{Page placement policy~$\pi$.}  The mapping from virtual pages to physical memory across partitions. Under round-robin interleaving at page granularity~$P$ across $N$ stacks, a contiguous virtual range of size~$F \gg P$ has approximately $\lceil F/P \rceil / N$ pages on any given partition.

  \item \textbf{WG scheduling policy~$\sigma$.}  The mapping of workgroup IDs to partitions.  Under round-robin dispatch, workgroup~$w$ executes on partition~$w \bmod N$.
\end{enumerate}
This decomposition separates the invariant kernel behavior (factor~1) from the policy-controlled runtime decisions (factors~2 and~3), making the metric applicable across placement strategies.

\paragraph{\textbf{Baseline locality under default policies}}
Under round-robin page interleaving and round-robin WG scheduling, as derived from AMD documents~\cite{AMD_CDNA3_Whitepaper, AMD_ROCm_Partitioning} and employed in our cycle-accurate simulator, any operand whose per-WG footprint~$F$ is large relative to the page size~$P$ converges to a baseline locality of
\begin{equation}
  \mathrm{Partition\;Locality_{baseline}} \;\approx\; \frac{1}{N},
  \label{eq:baseline-locality}
\end{equation}
regardless of sharing category.  On AMD Instinct MI300X with $N{=}4$~partitions, this baseline is approximately~$0.25$.  In other words, under naive
policies even WG-local operands suffer poor locality, because page interleaving spreads their footprint across all partitions indiscriminately.

\paragraph{\textbf{Achievable locality under optimized placement}}
What distinguishes the three sharing categories is not their baseline locality, but the \emph{ceiling} reachable through placement and scheduling optimization:
\begin{itemize}
  \item \textbf{WG-Private.}
    Each WG accesses a disjoint region of~$\mathcal{O}$.  The runtime can pin each WG's pages to its local partition memory, achieving 100\% locality with no data replication. This category offers the largest optimization headroom.

  \item \textbf{Global Sharing.}
    All WGs access the full extent of~$\mathcal{O}$.  Without replication, locality remains at~$1/N$ under any placement policy.  Achieving locality~$1.0$ requires $N$-way replication at a memory cost of~$N \times |\mathcal{O}|$.

  \item \textbf{Partial Sharing.}
    A subgroup of $G$~WGs shares a slice of~$\mathcal{O}$.  If all $G$~WGs are co-scheduled on the same partition and the shared slice is pinned to that partition's local memory, locality approaches~$1.0$.  Under default round-robin scheduling, the $G$~WGs spread across $\min(G, N)$~partitions, and locality degrades proportionally. This category benefits most from partition-aware WG scheduling.
\end{itemize}

\noindent
Thus, each sharing category implies a characteristic \emph{(baseline, ceiling)} pair: the baseline is what the default runtime delivers (${\approx}\,1/N$ in all cases), and the ceiling is what optimized placement and scheduling can achieve.  The gap between the two is the per-operand optimization headroom, and it is this gap that determines how much performance a NUMA-aware runtime can recover.

\paragraph{\textbf{Evaluation approach}}
In the next section, we report Partition Locality per operand for each target kernel under round-robin WG scheduling and page-interleaved placement at single-page granularity, matching the configuration of our cycle-accurate simulator (Section~\ref{sec:setup}).  The per-operand locality breakdown identifies which operand drives cross-partition traffic (the structural cause), while the simulator quantifies the latency overhead (the performance cost).

%% file: src/05_evaluation.tex
\section{Case-by-Case Analyses}
\label{sec:evaluation}


Having established the taxonomy of Section~\ref{sec:workload_characterization}, we now apply it across the target LLM operators. For each operator, we combine two complementary sources of evidence: memory trace analysis (Section~\ref{sec:methodology}), which identifies the sharing pattern and byte-level overlap of each operand from real kernel executions on the AMD Instinct MI300X, and cycle-level simulation, which quantifies the latency overhead incurred when participating workgroups are scheduled across partitions. Together, these measurements ground the taxonomy in hardware cost and allow us to evaluate operators by their NUMA sensitivity.

Each case study is chosen to stress a different point in the sharing-complexity spectrum: weight projection and MHA decode isolate the WG-private regime, GQA decode introduces partial sharing within head groups, MLA decode adds cross-kernel producer-consumer dependencies, FA prefill tests whether compute-boundness masks NUMA effects, and MoE exposes dynamic, routing-dependent sharing.


\subsection{Experimental Setup}
\label{sec:setup}

We run target kernels on an AMD Instinct MI300X under ROCm~7.1.0 for both memory trace collection and kernel object extraction. Table~\ref{tab:software} lists the software stack.
For memory trace generation we use Omniprobe~\cite{amdresearch_omniprobe}, an instrumentation tool for AMD Instinct GPUs that injects compile-time code to stream per-wavefront memory access messages.
Target GPU kernels are taken from vLLM and SGLang frameworks.

\begin{table}[t]
  \centering
  \caption{Software Environment}
  \label{tab:software}
  \small
  \begin{tabular}{ll}
    \toprule
    \textbf{Component}          & \textbf{Version / Details} \\
    \midrule
    GPU                          & AMD Instinct MI300X \\
    ROCm                         & 7.1.0 \\
    Trace tool                   & Omniprobe~\cite{amdresearch_omniprobe} \\
    Serving engines              & vLLM v0.0.19.2, SGLang v0.5.10 \\
    Cycle-level simulator        & MGPUSIM~\cite{sunMGPUSimEnablingMultiGPU2019} (extended) \\
    \bottomrule
  \end{tabular}
\end{table}



For performance analysis we extend MGPUSIM~\cite{sunMGPUSimEnablingMultiGPU2019},
a cycle-level GPU simulator, using the multi-partition
incorporation methodology of NVGIM~\cite{pratheekDesigningVirtualMemory2022} and
local/remote DRAM access latency information from
MCM-GPU~\cite{arunkumarMCMGPUMultiChipModuleGPUs2017} to model inter-partition memory
access overhead. We configure a 4-partition topology
similar to AMD Instinct MI300X~\cite{amdmi300x2025}. The remaining
simulation parameters are given in Table~\ref{tab:simconfig}
and are chosen to constitute a viable simulation setup.
Physical pages are distributed across partitions in round-robin
fashion at 4\,KB granularity, and workgroups are dispatched to
partitions in round-robin order at single-workgroup granularity
(WG $i$ executes on partition $i \bmod 4$), matching the
default policies documented by AMD~\cite{AMD_CDNA3_Whitepaper, AMD_ROCm_Partitioning}.

\begin{table}[b]
  \centering
  \caption{Simulated 4-Partition GPU Configuration}
  \label{tab:simconfig}
  \footnotesize
  \begin{tabular}{@{}ll@{}}
    \toprule
    \textbf{Parameter}          & \textbf{Value} \\
    \midrule
    Partitions (base dies)           & 4 \\
    CUs / partition             & 32 (8 SAs $\times$ 4 CUs) \\
    Total CUs                   & 128 \\
    Core frequency              & 2100\,MHz \\
    L1 cache                    & 32\,KB L1V/CU, 16\,KB L1S/4CUs\\
    L2 cache / partition        & 4\,MB \\
    L2 latency                  & 14-cycle bank + 2-cycle directory~\cite{pratheekDesigningVirtualMemory2022}\\
    DRAM / partition            & 2\,GB, 8 banks \\
    DRAM latency                & 95-cycle row-miss~\cite{arunkumarMCMGPUMultiChipModuleGPUs2017}\\
    Interconnect                & Crossbar, 32-cycle one-way, 768\,GB/s~\cite{arunkumarMCMGPUMultiChipModuleGPUs2017}\\
    Page placement              & 4\,KB round-robin~\cite{AMD_CDNA3_Whitepaper} \\
    WG scheduling               & Round-robin, per-WG granularity~\cite{AMD_ROCm_Partitioning} \\
    \bottomrule
  \end{tabular}
\end{table}


To isolate the performance cost of inter-partition data movement, we compare each kernel under two simulator configurations. The \emph{default} configuration uses the crossbar interconnect described in Table~\ref{tab:simconfig}. 
The \emph{ideal} configuration replaces the inter-partition interconnect with a zero-latency direct connection. Every inter-partition memory request completes in a single simulation cycle with no queuing, serialization, or bandwidth limitation, so that all accesses are served as if data were always partition-local. 

\begin{figure}[b]
  \centering
  \includegraphics[width=\columnwidth]{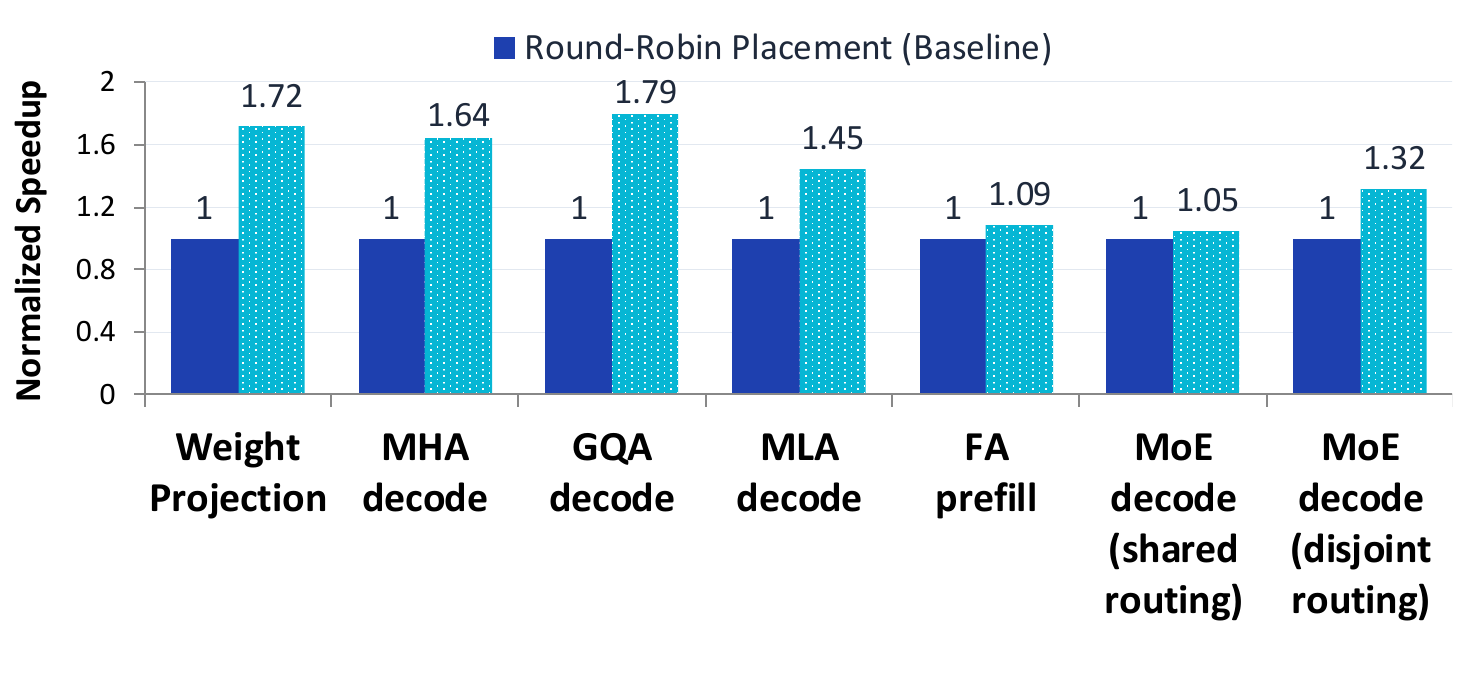}
  \caption{Achievable kernel speedup when inter-partition data transfer overhead is eliminated, normalized to default round-robin placement. Higher bars indicate greater NUMA sensitivity and larger optimization headroom.}
  \label{fig:latency}
\end{figure}

\subsection{Weight Projection}
\label{sec:results}
We analyze up-projection of Llama-2-7B~\cite{touvron2023llama2} with an FP16 $11008\times4096$ weight matrix in batch size 4. 
%
We use the TritonBLAS~\cite{swann2024tritonblas} backend, a Triton-based GEMM framework that analytically selects kernel configurations based on GPU cache hierarchy and memory topology via Origami~\cite{amd-origami-hipblaslt}, an AMD-developed model that predicts optimal tile sizes and grid strategies without autotuning.



\paragraph{\textbf{Trace observations}}
This kernel launches with a grid of $43\times1\times1$ workgroups.
Each WG reads a disjoint 2\,MB slice of the weight matrix in a WG-local pattern.
The input activation tensor (32\,KB across all four batch entries) is read in its entirety by every WG, constituting global shared access.
Each WG writes an independent 8\,KB output slice, again in a WG-local pattern.


\paragraph{\textbf{Partition Locality analysis}}
The weight matrix dominates the memory footprint at 2\,MB per WG, totaling $43 \times 2\,\text{MB} = 86\,\text{MB}$ across the full grid. 
Under default round-robin page interleaving at 4\,KB granularity, each WG's 2\,MB weight slice spans $512$ pages distributed across all $N{=}4$ partitions, yielding a baseline locality of $1/N = 0.25$.
Because weight access is WG-private, the achievable ceiling is 1.0 via per-WG page pinning with no data replication.

The input activation tensor (32\,KB) is globally shared but spans only 8 pages. Achieving locality 1.0 requires $N$-way replication at a cost of $4 \times 32\,\text{KB} = 128\,\text{KB}$, which is negligible.
Output tiles (8\,KB per WG) are WG-private and contribute minimally to cross-partition traffic.


\begin{figure}[h]
\centering
\includegraphics[width=\columnwidth]{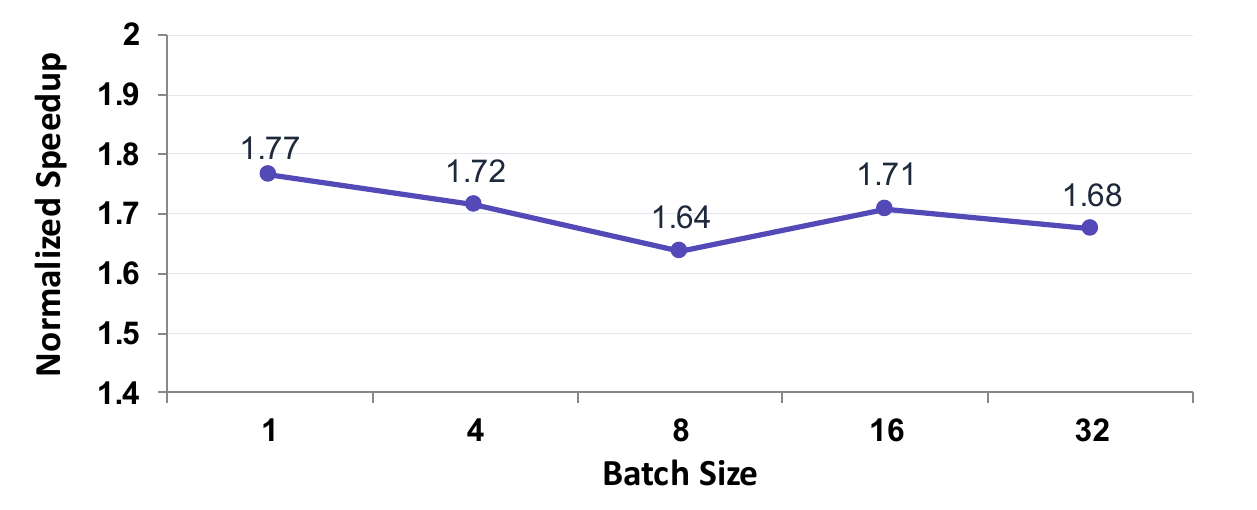}
\caption{NUMA sensitivity of weight projection across batch sizes. The dominant operand (2\,MB WG-private weight matrix) is batch-invariant. 
The globally shared activation grows from 8\,KB ($B{=}1$) to 256\,KB ($B{=}32$) but remains negligible relative to the weight footprint.
}
\label{fig:wproj-batch-sweep}
\end{figure}

\paragraph{\textbf{Performance impact}}
Decode-phase weight projection is a skinny matrix-matrix product ($11008 \times 4096 \times 4$), placing it squarely in the memory-bound regime where bandwidth utilization governs execution time.
Under default placement, approximately $3/4$ of the 2\,MB per-WG weight reads are served from remote partition memory, and the resulting inter-partition traffic directly translates to queuing delays and interconnect contention.
Cycle-level simulation comparing default round-robin placement against zero inter-partition overhead shows a 1.72$\times$ latency difference (Figure~\ref{fig:latency}).
Because all operands are either WG-private or negligibly small and globally shared, the optimization requires only per-WG page pinning with no WG co-scheduling, making weight projection the most amenable operator to static, partition-aware data placement among those we study.

In Figure~\ref{fig:wproj-batch-sweep}, we validate this across varying batch sizes.
The speedup remains within $1.64\text{--}1.77\times$ across all configurations, confirming that the NUMA sensitivity of weight projection is governed by the batch-invariant weight matrix rather than the scaling activation operand.

\subsection{Attention}
We analyze three attention variants that differ in KV head
structure and caching behavior: multi-head attention (MHA), which
maintains separate KV heads per query head; grouped-query attention
(GQA)~\cite{grattafiori2024llama3}, which shares each KV head across
a group of query heads; and multi-head latent attention
(MLA)~\cite{deepseekv2}, which compresses the KV cache into a
low-dimensional latent space. All implementations use
FlashAttention-based kernels~\cite{dao2024flashattention2}.

\subsubsection{Multi-Head Attention Decode}

For decode, we analyze the operation at batch size $B{=}4$ with KV cache of 1024 tokens previously generated from prefill.

\paragraph{\textbf{Trace observations}}
This kernel launches with a grid of $1\times32\times4$ workgroups, where grid Y dimension parallelizes heads and grid Z dimension parallelizes batch inputs. 
The per-token access granularity remains 256\,B, determined by the per-head dimension slicing ($d_{\text{head}} = 128$ elements in FP16).
Query vector, key and value caches are accessed in a WG-private pattern where each workgroup reads disjoint 256\,B slices of Q, K, and V with zero overlap as each WG handles a distinct (head, batch) pair and reads only the KV cache entries of its assigned sequence. 

\paragraph{\textbf{Partition Locality analysis}}
With all operands in the WG-private category, the achievable locality ceiling is $1.0$ where each WG's footprint can in principle be pinned to its executing partition's local memory.
However, under the default round-robin page-interleaved placement, the per-WG KV footprint of $1024 \times 256\,\text{B} = 256\,\text{KB}$ per operand is still distributed across all $N{=}4$ partitions, yielding a baseline locality of~${\approx}\,1/N = 0.25$.

\paragraph{\textbf{Performance impact}}
Cycle-level simulation shows a $1.64\times$ latency difference between default placement and ideal configurations.
Decode performs a single matrix-vector product per head rather than the matrix-matrix products of prefill, shifting the bottleneck from compute to memory bandwidth.
Under memory-bound regime, the $3/4$ fraction of remote partition memory accesses directly translates to execution time penalty.
Because all operands are WG-private, optimization requires pinning each WG's KV cache pages and query to the local memory of the partition that will execute that WG.
Similar to weight projection, MHA decode is an amenable target for simple per-WG data placement policies.

\subsubsection{Group Query Attention Decode}

GQA amortizes KV cache cost by sharing each KV head across a group of $G$ query heads. We analyze GQA decode of Llama-3-8B~\cite{grattafiori2024llama3} with $H_Q{=}32$ query heads and $H_{KV}{=}8$ KV heads ($G{=}4$), sequence length $L{=}1024$, head dimension $d_{\text{head}}{=}128$, and batch size $B{=}4$.

\paragraph{\textbf{Trace observations}}
GQA kernel has the same grid configuration as MHA, with a grid of $1\times32\times4$.
The per-token access granularity remains 256\,B ($d_{\text{head}} \times 2\,\text{B}$), consistent with MHA.
Query vector is read in WG-private pattern, identical to MHA decode.
On the other hand, Key and value caches are accessed in partial sharing pattern, per KV-head group. Each KV head is shared by $G{=}4$ query-head WGs. These four WGs read the same KV cache slice of $L \times d_{\text{head}} \times 2\,\text{B} = 256\,\text{KB}$ per operand, while WGs in different groups access disjoint KV slices.
This contrasts with MHA decode, where KV access is fully WG-private.
GQA introduces inter-WG sharing within each group of four, shifting KV from the private category to partial sharing.

\paragraph{\textbf{Partition Locality analysis}}
The total KV footprint is $H_{KV} \times 256\,\text{KB} \times 2 = 4\,\text{MB}$ (K and V combined), a $4\times$ reduction from MHA's $H_Q \times 256\,\text{KB} \times 2 = 16\,\text{MB}$.
Under default round-robin page interleaving, baseline KV locality remains~${\approx}\,1/N = 0.25$ per access, the same as MHA.
However, the achievable ceiling is no longer $1.0$ through simple per-WG pinning, because KV data is shared across $G{=}4$~WGs.
Reaching locality~$1.0$ for KV requires co-scheduling each group of 4~WGs onto the same partition and pinning the shared KV slice to that partition's local memory.
The query operand is accessed in WG-private pattern with achievable ceiling of~$1.0$ via per-WG pinning.

\paragraph{\textbf{Performance impact}}
Cycle-level simulation shows a $1.79\times$ latency difference between default placement and ideal configurations, compared to $1.64\times$ for MHA decode.
This higher overhead arises because GQA's grouped KV heads are accessed by multiple query-head workgroups simultaneously. With four query-heads sharing each KV head, WGs dispatched to different chiplets issue concurrent requests to the same remote cache lines, increasing inter-chiplet traffic by 23.7\% despite a 4$\times$ reduction in unique KV data, and raising average vector memory stall cycles per instruction from 1.96 to 2.44 compared to MHA.
%
%

\subsubsection{Multi-head Latent Attention Decode}
MLA~\cite{deepseekv2} compresses the KV
cache by storing a low-dimensional latent vector per token, reducing the per-token footprint from
$O(n_{\text{heads}} \cdot d_{\text{head}})$ to $O(d_{\text{latent}})$
at the cost of additional up-projection matrix multiplications during
decode to reconstruct K and V.
We preserve the core MLA dimensions, 512-dimensional
content, 64-dimensional RoPE, 576-dimensional keys, 512-dimensional values, and 128 query heads with sequence length of 1024 and batch size of 4. 

MLA decode differs from GQA and MHA in the compactness of the KV cache reads. Rather than reading per-head key and value vectors, each token contributes a single latent vector for K and a 512-dimensional latent vector for V, shared across all query heads. 
In our kernel, weight absorption is performed outside the attention kernel. Query projection pre-absorbs the up-projection weight matrix, so the MLA kernel reads only these latent vectors. 
The extreme 128:1 sharing ratio (all 128 Q-heads attending to one KV head) means every workgroup processing a different query head reads the identical KV sequence from cache, maximizing cache reuse but concentrating memory traffic on a narrow per-token footprint.
MLA decode of vLLM is a two-stage kernel pipeline. 
Stage 1 computes tiled attention over KV sequence splits, and Stage~2 reduces the partial outputs via log-sum-exp (LSE) trick analogous to the FlashDecoding formulation~\cite{dao2023flashdecoding}.

\paragraph{\textbf{Trace observations: Stage 1}}

This kernel launches with a $4\times8\times4$ grid (batch $\times$ query-head groups $\times$ sequence splits), where each WG operates on 16 query heads. 
Each workgroup reads 256 tokens worth of latent-dimension context, with each token occupying a 1024\,B block ($d_{\text{latent}} \times 2\,\text{B}$), yielding a per-WG K content read footprint of $256 \times 1024\,\text{B} = 256\,\text{KB}$. 
An additional 32\,KB of RoPE-dimension key is read per WG ($256 \times 64 \times 2\,\text{B}$). 
Value reads match the content dimension, contributing another 256\,KB ($256 \times 512 \times 2\,\text{B}$), for a total per-WG KV read footprint of 544\,KB. 
Both KV and query operands exhibit partial sharing: KV blocks are shared across WGs along grid axis Y, and query slices are shared along grid axis Z. Stage 1 writes intermediate results in 2052\,B granularity per block, a contiguous concatenation of 512 attention output elements and 1 log-sum-exp element, all in float32 precision ($512 \times 4 + 1 \times 4 = 2052\,\text{B}$).
Each WG writes 16 such blocks.
 
\paragraph{\textbf{Trace observations: Stage 2}}
This kernel launches with a grid of $4\times128\times1$ workgroups, where grid X dimension parallelizes batch inputs and grid Y dimension parallelizes reduction. 
Each Stage 2 workgroup reads four contiguous blocks of the (output~+~LSE) intermediate, totaling $4 \times 2052\,\text{B} = 8208\,\text{B}$ per WG in WG-private pattern. 
Each workgroup writes 1024\,B of attention output and a 4\,B LSE scalar, for a total per-WG write footprint of 1028\,B in a WG-private pattern.   
 
\paragraph{\textbf{Partition Locality analysis}}
Stage 1's sharing structure differs from MHA and GQA.
%
Latent KV (per-WG footprint of 544\,KB) is shared across all WGs across grid Y, forming subgroups whose size depends on the grid Y dimension. Under default round-robin page interleaving, baseline locality is~${\approx}\,1/N = 0.25$. The achievable ceiling requires co-scheduling each Y-group onto the same partition and pinning the corresponding latent slice locally.
Query is partial shared across grid Z, forming a second, orthogonal partial sharing group.
Stage 1 intermediate output is a WG-private write where each WG writes its own 16 blocks of $2052\,\text{B}$ with no inter-WG overlap.
 

\paragraph{\textbf{A Case for Inter-kernel Data Reuse}}
The two-stage formulation reveals an important case of inter-kernel data sharing.
Stage 1 writes intermediate operands in a WG-private pattern, and Stage 2 reads these operands in the same WG-private category.
However, each Stage 1 WG writes 16 intermediate blocks, while each Stage 2 WG reads four contiguous blocks, each produced by different Stage 1 WGs.
This creates a producer-consumer dependency across kernel boundaries where a single Stage 2 consumer depends on four distinct Stage 1 producers.
Since the intermediate buffer is small enough to reside in L2 cache, this is a case where partition-aware WG scheduling can exploit L2 reuse across kernel boundaries, provided the four Stage 1 producers and their corresponding Stage 2 consumer are co-scheduled on the same partition.
%
This demonstrates that NUMA-aware scheduling must consider not only intra-kernel sharing but also cross-kernel relationships.

\paragraph{\textbf{Performance impact}}
Cycle-level simulation shows a $1.45\times$ latency difference between default placement and ideal configurations.
This is lower than both MHA decode ($1.64\times$) and GQA decode ($1.79\times$), because the per-token KV cache footprint is substantially smaller than the $n_{\text{heads}} \times d_{\text{head}} \times 2\,\text{B}$ of MHA or GQA, reducing total cross-partition traffic.
However the scheduling complexity is higher. Optimizing MLA locality demands coordinating both intra-kernel group affinity across two grid axes and inter-kernel producer-consumer co-location, whereas MHA requires only per-WG pinning.

\begin{figure}[h]
\centering
\includegraphics[width=\columnwidth]{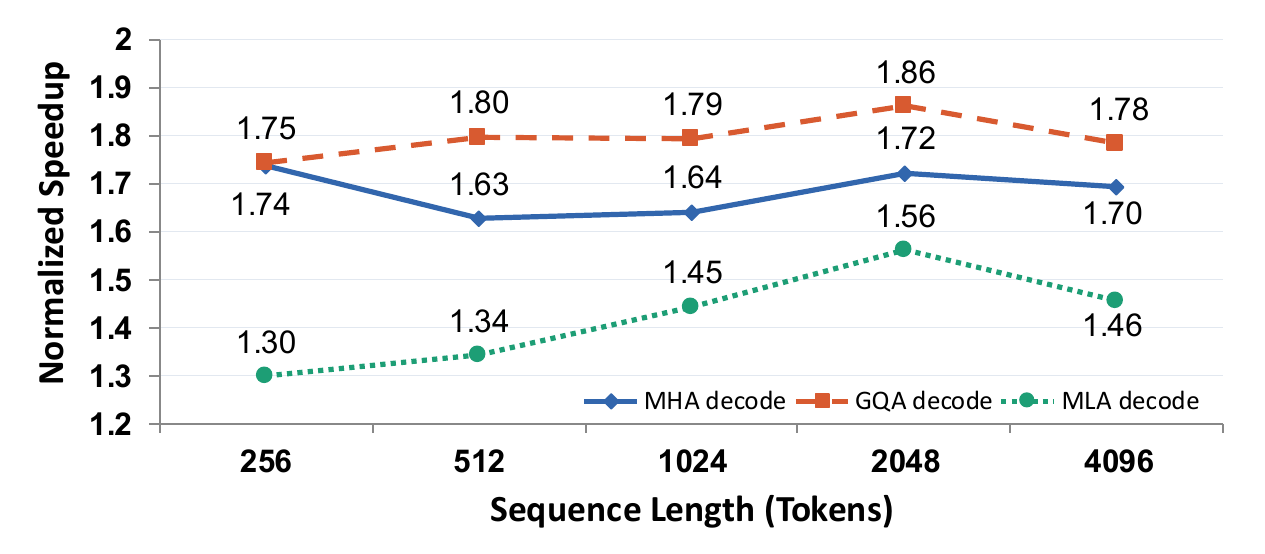}
\caption{NUMA sensitivity of attention decode variants across
sequence lengths at batch size $B{=}4$. The sharing category of
each operand is stable across all lengths 
, while
the overhead magnitude grows with per-WG KV footprint.}
\label{fig:attn-seqlen-sweep}
\end{figure}

In Figure~\ref{fig:attn-seqlen-sweep}, we validate the NUMA sensitivity of all three decode variants across varying sequence lengths 
at $B{=}4$. GQA decode consistently exhibits the highest overhead
($1.75\text{--}1.86\times$) due to contention amplification from
partial KV sharing, followed by MHA ($1.56\text{--}1.74\times$)
and MLA ($1.30\text{--}1.72\times$), confirming that the sharing
category and NUMA sensitivity are preserved
across sequence lengths.

\subsubsection{Flash Attention Prefill}
We profile MHA FlashAttention prefill at single batch, sequence of length $L{=}1024$, with $H{=}32$ heads and hidden dimension $d_{\text{model}}{=}4096$.


\paragraph{\textbf{Trace observations}}
This kernel launches with a $8\times32\times1$ grid, where grid X dimension parallelizes query heads and grid Y dimension splits the sequence length.
Each head operates on a per-head dimension of $d_{\text{head}} = 128$ elements in FP16, yielding a per-token access granularity of $128 \times 2\,\text{B} = 256\,\text{B}$ for each of the query, key, and value operands.
%
Each workgroup reads its own query tile, consisting of one or more contiguous 256\,B blocks corresponding to its assigned output rows. 
Key and Value activations exhibit partial sharing. All workgroups assigned to the same attention head read the \emph{entire} key and value sequences for that head. Each WG iterates over all $L$ tokens during the tiled score computation and value aggregation. 

\paragraph{\textbf{Partition Locality analysis}}
While the per-token access granularity is 256\,B, the relevant quantity for Partition Locality is the aggregate per-head footprint, the total KV data that a head's WG subgroup collectively touches.
For sequence length $L{=}1024$, this is $256\,\text{KB}$
%
%
per operand (K or V), totaling 512\,KB of shared KV data per head.
At $P{=}4\,\text{KB}$ interleaving, the 256\,KB per-operand footprint spans ${\sim}64$~pages distributed round-robin across $N{=}4$~partitions, and baseline locality converges to~${\approx}\,1/N = 0.25$.
%
As query access is WG-private, the achievable ceiling under optimized placement is~$1.0$ with no replication cost.


\paragraph{\textbf{Performance impact}}
While the partial sharing structure of KV within a head makes \emph{head-to-partition affinity} the primary optimization direction, co-scheduling all WGs of the same attention head onto a single partition, and pinning that head's KV activations to the same partition's local memory to raise KV locality from the baseline toward~$1.0$, cycle-level simulation proves this unnecessary.

Cycle-level simulation comparing default round-robin placement against the ideal configuration shows only a $1.09\times$ latency difference.
This is consistent with the compute-bound nature of FlashAttention prefill. The tiled matrix multiplications in the score computation and value aggregation dominate execution time, and the memory subsystem (including remote partition memory accesses) is not on the critical path.
Although KV locality is poor under default placement, the arithmetic intensity of prefill masks the NUMA penalty.
This stands in contrast to the decode-phase attention kernels, where memory-boundness exposes the same locality deficiency as a significant cost.




\subsubsection{Mixture-of-Experts}
We evaluate the fused MoE kernel from vLLM using DeepSeek-V3~671B dimensions: hidden dimension 7168, expert dimension 2048, 256
total experts with top-$K{=}8$ routing. 
Considering a single-GPU scenario under expert parallelism, we use 32 experts with 8 activated, with batch size 4.
The kernel implements a batched GEMM across all active experts in a single dispatch. 
A pre-processing step sorts tokens by their routed expert and pads to \texttt{BLOCK\_SIZE\_M}{=}64 boundaries, producing a contiguous token index array and a per-block expert assignment map.
Each WG is assigned a $(\mathit{pid\_m}, \mathit{pid\_n})$ tile: $\mathit{pid\_m}$ determines the expert and token block, $\mathit{pid\_n}$ determines the output feature slice. 
This mapping is central to the sharing analysis as it defines which WGs read overlapping activation rows.

\paragraph{\textbf{Trace observations}}
Each WG reads three operand classes. Routing metadata 
is globally shared, while
input activations are partially shared. 
Each token is routed to $K{=}8$ experts, each with 32 output-tiling WGs,
so $8 \times 32 = 256$ WGs read the same activation rows. In the
decode setting with four real tokens per expert block, the actual
activation access per WG is
$4 \times 7168 \times 2\,\text{B} = 56\,\text{KB}$.
Expert weights ($64 \times 7168 \times 2\,\text{B} = 896\,\text{KB}$ per WG) are
WG-private. 
Each WG reads a distinct \textit{BLOCK\_SIZE\_N}-wide strip of its expert's weight matrix, with no overlap across output tiles.
Output tiles ($8\,\text{KB}$ per WG) are WG-private.
 
\paragraph{\textbf{Partition Locality analysis}}
The dominant cross-partition traffic sources are expert weights
(896\,KB per WG, WG-private) and input activations (56\,KB per WG
at $B{=}4$, partial sharing with group size $K{=}8$). Routing
metadata (${\sim}256$\,B) and output tiles (8\,KB) are too small
to contribute meaningfully. Under default page interleaving, both
dominant operands converge to baseline locality ${\approx}\,1/N$,
but they require different optimizations. Weight locality can be
raised to $1.0$ via per-WG page pinning, whereas activation
locality requires co-scheduling the $K{=}8$ expert WGs that share
token rows onto the same partition. Expert weights contribute a
fixed 896\,KB per WG regardless of batch size, while activation
footprint scales linearly with the number of routed tokens per
expert block, making activation placement equally critical at
larger batch sizes.

 
\paragraph{\textbf{Routing-dependent sharing and placement sensitivity}}
Because the kernel dispatches all active experts in a single grid, the
token-to-expert routing outcome determines both the total unique weight
footprint and the degree of cross-WG reuse. We evaluate two extremes.
Under \emph{shared routing}, all four tokens activate experts 0--7 and thus the
unique weight working set is $8 \times 2048 \times 7168 \times 2\,
\text{B} = 224$\,MB, and each expert's WGs read all four tokens. 
Under \emph{disjoint routing}, each token activates eight entirely
different experts ($4 \times 8 = 32$), so every expert handles a
single token. The unique weight footprint grows to
$896$\,MB with zero cross-WG reuse, pressuring the cache
hierarchy and compounding the interconnect overhead under default
page interleaving.
 
Critically, token routing dynamically determines both the number of active WGs and the total expert weight footprint. 
Without adequate co-placement of WGs and their corresponding data, this variability leads to repeated cross-partition reads of both expert weights and activations.

\begin{table*}[t]
\centering
\caption{Per-operand sharing patterns, footprints, and optimization strategies across LLM kernels.
Speedup indicates the latency ratio between ideal (zero inter-partition overhead) and default
round-robin placement at $B{=}4$, $L{=}1024$ unless noted.
Negligible operands (output tiles $\leq$8\,KB, routing metadata $\sim$256\,B) are omitted.}
\label{tab:summary}
\small
\begin{tabular}{@{}clccccl@{}}
\toprule
\textbf{Kernel} & \textbf{Operand} & \textbf{Sharing} & \textbf{Per-WG} & \textbf{Subgroup} & \textbf{Speedup} & \textbf{Required Optimization} \\
 &  & \textbf{Pattern} & \textbf{Footprint} & \textbf{Size} & & \\
\midrule
 
\multirow{2}{*}{\makecell[c]{Weight\\Projection}}
  & Weight matrix   & Private & 2\,MB   & ---       & \multirow{2}{*}{1.72$\times$} & Per-WG page pinning \\
  & Activation      & Global  & 32\,KB  & All WGs   &                                & $N$-way replication (128\,KB) \\
 
\midrule
 
\multirow{2}{*}{\makecell[c]{MHA\\Decode}}
  & Query           & Private & 256\,B  & ---       & \multirow{2}{*}{1.64$\times$} & Per-WG page pinning \\
  & KV cache        & Private & 256\,KB & ---       &                                & Per-WG page pinning \\
 
\midrule
 
\multirow{2}{*}{\makecell[c]{GQA\\Decode}} 
  & Query           & Private & 256\,B  & ---       & \multirow{2}{*}{1.79$\times$} & Per-WG page pinning \\
  & KV cache        & Partial & 256\,KB & $G{=}4$   &                                & Co-schedule $G$ WGs + pin \\
 
\midrule
 
\multirow{3}{*}{\makecell[c]{MLA\\Decode}} 
  & Latent KV       & Partial & 544\,KB & $|Y|{=}8$ & \multirow{3}{*}{1.45$\times$} & Co-schedule Y-group + pin \\
  & Query           & Partial & 256\,B  & $|Z|{=}4$ &                                & Co-schedule Z-group \\
  & Intermediate    & Private & 32\,KB  & ---       &                                & Inter-kernel co-location\textsuperscript{\textdagger} \\
 
\midrule
 
\multirow{2}{*}{\makecell[c]{FA\\Prefill}} 
  & Query           & Private & 256\,B/tile & ---        & \multirow{2}{*}{1.09$\times$} & Per-WG page pinning \\
  & KV              & Partial & 256\,KB/head        & Per head   &                                & Head-to-partition affinity \\
 
\midrule
 
\multirow{2}{*}{\makecell[c]{MoE\\Decode}} 
  & Expert weights  & Private & 896\,KB & ---       & \multirow{2}{*}{\makecell{1.05$\times$\textsuperscript{S}\\1.32$\times$\textsuperscript{D}}} & Static per-WG pinning \\
  & Activation      & Partial & 56\,KB\textsuperscript{\textsection}  & $K{=}8$   &                                & Dynamic per-batch co-schedule \\
 
\bottomrule
\end{tabular}
 
\vspace{2pt}
{\footnotesize 
\textsuperscript{\textdagger}Cross-kernel producer–consumer locality between Stage 1 and Stage 2 enables L2 reuse under partition co-location.\\
\textsuperscript{S}Shared routing (all tokens $\to$ same 8 experts). \textsuperscript{D}Disjoint routing (each token $\to$ different 8 experts).\\
\textsuperscript{\textsection}At $B{=}4$ (4 real tokens per expert block); scales linearly with batch size.
}
\end{table*}

\paragraph{\textbf{Performance impact}}
Cycle-level simulation confirms this sensitivity. The latency gap
between default and ideal configurations is $1.05\times$ under shared
routing but $1.32\times$ under disjoint routing, reflecting the
larger unique weight footprint and absence of cross-WG reuse.

Closing this gap requires a two-tier placement strategy. A
\emph{static} tier pins each expert's weight tiles and its
corresponding WGs to the same partition. A \emph{dynamic} tier places
token activation pages on the partitions whose experts were selected
by the router, which is a decision that changes every batch. Neither tier
alone suffices as pinning weights without co-locating activations still
incurs remote activation reads, while co-locating activations without
pinning weights leaves the dominant operand scattered across
partitions.



\subsection{Discussion}
\label{sec:discussion}
Table~\ref{tab:summary} consolidates the sharing patterns,
per-workgroup footprints, achievable speedups, and required
placement optimizations identified across the analyzed
LLM kernels.
Throughout different case studies on LLM kernels on multi-partition GPU, we identified memory trace-based WG-level data access patterns and derived achievable Partition Locality under optimal workgroup/data placement. With a cycle-level simulator, we derived the achievable kernel speedup by optimizing inter-partition data access.

Returning to our motivating question, the cost of NUMA in LLM serving kernels ranges from negligible (1.09$\times$ for compute-bound FA prefill) to substantial (1.79$\times$ for GQA decode), and the required mitigation ranges from simple per-WG page pinning (weight projection, MHA decode) to subgroup-aware co-scheduling with dynamic, per-batch placement (MoE, MLA). Crucially, no single placement policy suffices across all operators. The sharing category of the dominant operand determines both the severity of the penalty and the complexity of the solution. However, achieving this optimal placement requires significant support from the underlying GPU architecture.


\subsubsection{GPU Architecture Implications}

\paragraph{\textbf{Data Interleaving Granularity}}
Our case studies show that the per-WG operand slices vary widely in size across kernels from 8\,KB output tiles to 2\,MB weight slices in projection, and from 256\,B query vectors to 544\,KB latent KV blocks in MLA decode.
Under GPU's fixed 4\,KB page-size interleaving, this variation complicates correct chiplet-local placement of data structure slices for each WG.

CLAP~\cite{parkLeveragingChipletLocalityEfficient2025} samples a subset of pages at runtime to profile each data structure's partition-locality degree, then maps page groups contiguously into the physical frames of the predicted home partition. The resulting virtual-to-physical contiguity allows these groups to be covered by a single merged TLB entry, delivering large-page TLB reach without the coarse-grained misplacement that large pages impose in NUMA settings. However, CLAP assumes round-robin WG scheduling whereas our partial-sharing results (GQA, MoE) show that data placement alone is insufficient without coordinated WG co-scheduling.

\paragraph{\textbf{WG Scheduling Granularity}}
For WG-private operands, it is crucial to map each WG--data structure slice pair to the same partition.
For partially shared operands, the relevant co-location unit is the WG subgroup. In GQA decode, the $G{=}4$ query-head WGs sharing a KV head must be co-scheduled. In MoE, expert-specific WGs sharing routed activations must land together. 
These complex WG subgroup--data structure slice dependencies call for explicit, fine-grained placement of WGs across partitions, beyond default round-robin dispatch.

LADM (Locality-Aware Data Management)~\cite{khairy_locality-centric_2020} introduces a compiler-assisted static index analysis that classifies GPU kernel access patterns into three categories, no datablock-locality, row/column locality, and intra-thread locality, and uses this to have GPU driver proactively co-place threadblocks (or workgroups) and their data onto the same partition before kernel execution, avoiding the latency penalties of reactive first-touch demand paging.
MGvm (MCM-aware GPU virtual memory)~\cite{pratheekDesigningVirtualMemory2022} builds on LADM's static analysis for workgroup scheduling and data placement, and additionally extends it to the virtual memory layer. It coordinates TLB home-slice assignment with PTE placement to keep address translation local, accompanied by a 2MB--4KB interleaving granularity switching mechanism driven by a real-time hardware monitor to prevent L2 TLB contention hotspots. While MGvm addresses WG-private patterns effectively, the dynamic per-batch placement required by MoE's routing-dependent sharing (Section 5.3.5) remains outside its static analysis scope.


\subsubsection{Kernel Programming Implications}
The architectural solutions discussed above operate below the programming model. Page placement policies and WG scheduling decisions are made by the MMU and hardware dispatcher, neither of which is exposed to kernel code.
From a kernel developer's perspective, the physical placement of data across partition memory is entirely
invisible. There is no programmatic mechanism to pin an operand to a specific partition's memory. 
What kernel developers \emph{can} control is the mapping of workgroups to compute resources, which in turn determines which WGs
share an on-die L2 cache. This makes WG scheduling order the primary lever available at the kernel programming level for improving data locality on multi-partition GPUs.

For example, Choudhary et al.~\cite{choudharyOptimizingAttentionGPUs2025}  demonstrate that by swizzling the workgroups in head-first mapping, they can localize workgroups of a head that share K and V tensors to the same XCD, confining shared data to a single per-die L2 cache and achieving up to 50\% throughput improvement over conventional scheduling with L2 hit rates of 80--97\%.

AMD's CPX (Core Partitioned X-celerator) mode~\cite{AMD_ROCm_Partitioning} offers a more direct alternative. Combined with NPS4 memory partitioning, CPX exposes each XCD as a separate logical GPU with its own local memory partition, giving kernel
developers explicit control over both compute and data placement. This sidesteps the abstraction gap entirely, but at the cost of the unified address space. Each partition sees only a fraction of total memory, and cross-partition communication requires explicit multi-GPU coordination.